\documentclass[aps,twocolumn,showpacs,floats,prl]{revtex4}
\usepackage{graphicx}
\usepackage{dcolumn}
\usepackage{bm}
\usepackage{color}
\usepackage{amsmath}
\usepackage{subfigure}
\usepackage{enumitem}
\usepackage{ dsfont }
\usepackage{amssymb}

\begin{document}

\author{L. A. Pe\~{n}a Ardila, Silvio A. Vitiello	 and Maurice de Koning $^{*}$}
\affiliation{Instituto de F\'isica Gleb Wataghin, Universidade Estadual de Campinas, UNICAMP, 13083-859, Campinas, S\~ao Paulo, Brazil}

\title{Elastic constants of hcp $^{4}$He: Path-integral Monte Carlo results versus experiment} 

\begin{abstract} 
The elastic constants of hcp $^{4}$He are computed using the path-integral Monte Carlo (PIMC) method.  The stiffness  coefficients  are  obtained  by  imposing  different  distortions  to  a  periodic  cell  containing  180  atoms, followed by measurement of the elements of the corresponding stress tensor. For this purpose an appropriate path-integral expression for the stress tensor observable is derived and implemented into the PIMC++ package. In addition to allowing the determination of the elastic stiffness constants, this development also opens the way to an explicit atomistic determination of the Peierls stress for dislocation motion using the PIMC technique. A  comparison  of  the  results  to  available  experimental  data  shows  an  overall  good  agreement  of  the  density dependence of the elastic constants, with the single exception of C$_{13}$. Additional calculations for the bcc phase, on the other hand, show good agreement for \textit{all} elastic constants.
\end{abstract}


\maketitle

\section{I. Introduction}
\label{sec:intro}

Despite the intensive research efforts developed over the past six years, the remarkable results of the torsional-oscillator (TO) experiments on solid $^{4}$He by Kim and Chan~\cite{1,2} still elude a consistent explanation. Although it has been suggested that the observed nonclassical rotational inertia (NCRI) is a manifestation of superfluidity in the solid phase,~\cite{3,4,5} several questions remain unanswered. One of them is the apparent correlation between the NCRI data and the observation of elastic stiffening upon cooling, both of which display similar temperature and $^{3}$He-concentration  dependences~\cite{6,7}. The  observed  stiffening effect  has  been  interpreted  in  terms  of  dislocation  pinning due to  $^{3}$He impurities~\cite{6,8} and this has contributed to ideas that the  NCRI  may  not  have  an  exclusive  non-superfluid  origin  but rather that it might also be a manifestation of mechanical behaviour~\cite{8,9,10,11,12}.

A major difficulty is the fact that the insight obtained from
recent experiments~\cite{6,8,13} relies on the indirect interpretation of
data, a process that is inevitably based on assumptions. The
dislocation-pinning interpretation, for instance, conjectures
the presence of dislocation networks that are pinned by $^{3}$He
impurities binding to the dislocation cores~\cite{6,8}. Despite the
elevated degree of sophistication of recent experiments~\cite{6,8,13},
however, it has not yet been possible to verify these assumptions
explicitly, hampering a conclusive understanding of the
observed phenomenology and its relation to the NCRI data
obtained in TO experiments.

In this context, theoretical modelling can serve as a useful
complementary approach. In principle, it allows a systematic
study of a hierarchy of well-controlled structures not
accessible to experiment, starting at the defect-free crystal
and progressing through a series of configurations containing
different defect geometries. Analysis of the results obtained in
these different situations may then assist in the interpretation
of existing experimental data or even lead to new predictions.
However, a first step in this effort is to gauge the results of
these modelling approaches for quantities that can be directly
compared to experimental data. In the context of the recent
observations of the mechanical behaviour of solid $^{4}$He, its
intrinsic elastic properties are an evident target for such an
assessment.

The purpose of the present paper is to compare the results
of state-of-the-art path-integral Monte Carlo~\cite{14,15} (PIMC)
calculations based on the Aziz pair potential~\cite{16} to the available
experimental data for the elastic stiffness constants from the
1970s. The PIMC method has shown to give excellent agreement
with experiment for several properties of the liquid and
solid states, including the energy, pair correlation functions,
structure factors, and superfluid density. Triggered by TO
experiments of Kim and Chan, the PIMC methodology has
also been applied extensively to the study of $^{4}$He supersolidity,
focusing on a variety of properties, including the possibility
of superflow induced by lattice defects such as vacancies~\cite{17,18}
and dislocations~\cite{19,20}.

The first path-integral approach to the computation of
elastic constants was based on a direct measurement of the
elastic constants in terms of the second derivatives of the
partition function with respect to strain components~\cite{21}. Here
we adopt a different approach based on the development of
an expression for the stress-tensor observable in the path integral
formalism. Not only does this observable allow the
determination of the elastic constants of the defect-free crystal,
it is also a key observable in the characterization of plastic
deformation in terms of lattice defect properties. In this light,
this development is of interest in its own right, allowing for instance,
an explicit atomistic determination of the Peierls stress
for dislocation motion~\cite{22}. Here we utilize it to compute the
complete set of stress-strain relations by measuring the elastic
stress response to small homogeneous deformations, giving
the five independent elastic stiffness constants of the hcp $^{4}$He
crystal. In addition to a direct comparison of elastic constants
as a function of density, we analyze the degree to which the
Cauchy relation, which measures to what extent non central
forces and zero-point effects are important, is satisfied.

The remainder of the paper has been organized as follows.
In Sec. II we derive an expression for the stress-tensor
observable within the path-integral formalism based on the
pair-action approximation and which has been implemented
in the PIMC++ code~\cite{18}. Section III describes the employed
computational setup and summarizes the parameters used in
the simulations. Next, we describe and discuss the obtained
results in Sec. IV and summarize in Sec. V.

\section{II. PATH-INTEGRAL EXPRESSION FOR THE STRESS-TENSOR OBSERVABLE.}
\label{sec:iPATH-INTEGRAL EXPRESSION FOR THE
STRESS-TENSOR OBSERVABLE}

Following the usual approach of Parrinello and
Rahman~\cite{23,24,25}, we describe homogeneous distortions of a
periodic simulation cell in terms of the box matrix h, whose
columns are the three periodic repeat vectors a, b, and c of the
computational cell~\cite{25},

\begin{equation}
\mathds{h}=\left(\begin{array}{ccc}
a_{x} & b_{x} & c_{x}\\
a_{y} & b_{y} & c_{y}\\
a_{z} & b_{z} & c_{z}
\end{array}\right)
\label{eq:1}
\end{equation}

In terms of the $\mathds{h}$ matrix, an absolute position $r_{i}$ within in the
cell is written as
\begin{equation}
r_{i}=\mathds{h} s_{i,}
\label{eq:2}
\end{equation}

where $s_{i}$ is a relative coordinate vector whose components
have values ranging between 0 and 1. In this description, homogeneous
deformations are described in terms of variations
of the $c$ matrix at fixed relative coordinates $s_{i}$.

~To obtain an expression for the stress tensor in the path integral
formalism we start with its thermodynamic definition,
describing its components $\sigma_{ij}$ in terms of derivatives of the
appropriate thermodynamic potential. Specifically, we have~\cite{25}

\begin{equation}
\sigma_{ij}=-\frac{1}{\mathtt{\det(\mathds{h})}}\sum_{k=1}^{3}\mathds{h}_{jk}\left(\frac{\partial F}{\partial \mathds{h}_{ik}}\right)_{N,T},
\label{eq:3}
\end{equation}
where
\begin{equation}
F=F(N,\mathds{h},T)
\label{eq:4}
\end{equation}
is the Helmholtz free energy of a system containing $N$ particles
that are confined to a volume described by the matrix $\mathds{h}$ and
in thermal equilibrium with a heat bath at temperature $T$.
As usual, the microscopic description for the stress tensor is
obtained through the connection between the thermodynamic
potential and the corresponding partition function. Here, we
have
\begin{equation}
F(N,\mathds{h},T)=-\frac{1}{\beta}\ln Z(N,\mathds{h},T)
\label{eq:5}
\end{equation}

where $Z(N,\mathds{h},T)$ is the canonical partition function and $\beta=(K_{B}T)^{-1}$. In this manner, the components of the stress tensor $\mathds{P}$ become

\begin{equation}
\mathds{P}_{ij}=-\frac{1}{\mathtt{\det(\mathds{h})\beta}Z}\sum_{k=1}^{3}\mathds{h}_{jk}\left(\frac{\partial Z}{\partial \mathds{h}_{ik}}\right)_{N,T},
 \label{eq:6}
\end{equation}

~In the path-integral formalism, the canonical partition
function for a system of distinguishable particles can be written
as~\cite{14}

\begin{equation}
\begin{split}
Z(N,\mathds{h},T)=\int\cdots\int dR_{0}\cdots dR_{M-1}\\
\times \exp\left[-S_{\mathtt{path}}(R_{0,\cdots,}R_{M-1},R_{0})\right]
 \label{eq:7}
\end{split}
\end{equation}

where $R_{k}=\{r_{1,k},\cdots,r_{N,k}\}$ represents the set of position vectors
of the $N$ particles in the $k$th time slice of theM-bead closed
path $R=\{R_{0,\cdots,}R_{M-1},R_{0}\}$, and $S_{\mathtt{path}}(R_{0,\cdots,}R_{M-1},R_{0})$
is the path action. Proper symmetrization for the case of
indistinguishable particles is straightforward~\cite{14}.
~Describing the position vectors in terms of the $\mathds{h}$ matrix,
the expression becomes

\begin{equation}
\begin{split}
Z(N,\mathds{h},T)=\int\cdots\int\left[\det(\mathds{h})\right]^{NM}dS_{0}\cdots dS_{M-1}\\
\times \exp\left[-S_{\mathtt{path}}(S_{0,\cdots,}S_{M-1},S_{0};\mathds{h})\right],
\label{eq:8}
\end{split}
\end{equation}

where $S_{k}=\{s_{1,k},\cdots,s_{N,k}\}$ represents the set of scaled position
vectors of the $N$ particles in the $k-$th time slice. The
derivatives of the partition function with respect to the elements
of the $\mathds{h}$ matrix are then given by

\begin{equation}
\begin{split}
\left(\frac{\partial Z}{\partial \mathds{h}_{ij}}\right)_{N,T}=\left[\det(\mathds{h})\right]^{NM}\int\cdots\int dS_{0}\cdots dS_{M-1}\\
\times \exp\left[-S_{\mathtt{path}}\right]\left[NM(H)_{ji}^{-1}-\frac{\partial S_{\mathtt{path}}}{\partial \mathds{h}_{ij}}\right].
\label{eq:9}
\end{split}
\end{equation}

Substitution into Eq.~(\ref{eq:6})  then gives

\begin{equation}
\mathds{P}_{ij}=\frac{1}{\mathtt{\det(\mathds{h})\beta}}\left[NM\delta_{ij}-\sum_{k=1}^{3}\mathds{h}_{jk}\left\langle \frac{\partial S_{\mathtt{path}}}{\partial \mathds{h}_{ik}}\right\rangle \right]
 \label{eq:10}
\end{equation}

where the angular brackets indicate averaging over closed
paths. Given that the paths consist of $M$ links, the above
expression can also be written in terms of averages over link
actions, namely,

\begin{equation}
\mathds{P}_{ij}=-\frac{M}{\mathtt{\det(\mathds{h})\beta}}\left[N\delta_{ij}-\sum_{k=1}^{3}\mathds{h}_{jk}\left\langle \frac{\partial S_{\mathtt{Link}}}{\partial \mathds{h}_{ik}}\right\rangle \right]
 \label{eq:11}
\end{equation}

The hydrostatic pressure $P$ is then given by

\begin{equation}
P=\frac{1}{3}\mathtt{Tr}\mathds{P}=-\frac{M}{\mathtt{\det(\mathds{h})\beta}}\left[N-\frac{1}{3}\mathds{h}_{ik}\left\langle \frac{\partial S_{\mathtt{Link}}}{\partial \mathds{h}_{ik}}\right\rangle \right].
 \label{eq:12}
\end{equation}

Next we determine the derivative of the link action with
respect to the elements of the $\mathds{h}$ matrix. Here, we are
specifically interested in the pair approximation for the action,

\begin{equation}
\begin{split}
S_{link}=S(R,R';\tau)=S_{kin}(R,R';\tau)+S_{pot}(R,R';\tau)\\
=\sum_{n=1}^{N}K(r_{n},r'_{n};\tau)+\sum_{n<m}u_{2}(r_{n}-r_{m},r'_{n}-r'_{m};\tau),
 \label{eq:13}
\end{split}
\end{equation}

~where the first term represents the exact kinetic link action,
the second is potential action within the pair-product
approximation~\cite{14} and $\tau=\beta/M$. It is important to emphasize
that, while referred to as the potential action, $u_{2}$ in fact contains
the remainder of the \textit{exact} two-body action (i.e., including both
kinetic and potential parts) after separating out the kinetic link
action $K$~\cite{14}.
The derivatives of interest are then

\begin{equation}
\begin{split}
\frac{\partial S_{\mathtt{Link}}}{\partial \mathds{h}_{ij}}=&\sum_{n=1}^{N}\frac{\partial K(r_{n},r'_{n};\tau)}{\partial \mathds{h}_{ij}}\\
&+\sum_{n<m}\frac{\partial u_{2}(r_{n}-r_{m},r'_{n}-r'_{m};\tau)}{\partial \mathds{h}_{ij}}.
 \label{eq:14}
\end{split}
\end{equation}

The kinetic action is given by

\begin{equation}
\begin{split}
K(r,r';\tau)=\frac{3}{2}\ln\left(4\pi\lambda\tau\right)+\frac{\left|r-r'\right|^{2}}{4\lambda\tau},
 \label{eq:15}
\end{split}
\end{equation}

with $\lambda=\hbar^{2}/2m$, and is normalized such that 

\begin{equation}
\begin{split}
\int dr'\exp\left[-K(r,r';\tau)\right]=1.
 \label{eq:16}
\end{split}
\end{equation}

To compute the derivatives with respect to the elements of the
$\mathds{h}$ matrix in Eq.~(\ref{eq:14}) , we write Eq.~(\ref{eq:15})  in terms of $\mathds{h}$ and the
relative coordinates $s$. The result is

\begin{equation}
\begin{split}
K(s,s';\mathds{h},\tau)=\frac{3}{2}\ln\left(4\pi\lambda\tau\right)+\ln\left[\det(\mathds{h}^{-1})\right]+\frac{\left|\mathds{h}(s-s')\right|^{2}}{4\lambda\tau},
 \label{eq:17}
\end{split}
\end{equation}

which is normalized such that

\begin{equation}
\begin{split}
\int ds'\exp\left[-K(s,s';\mathds{h},\tau)\right]=1.
 \label{eq:18}
\end{split}
\end{equation}

The derivatives with respect to the $\mathds{h}$-matrix elements are
then given by

\begin{equation}
\begin{split}
\frac{\partial K(s,s';\mathds{h},\tau)}{\partial \mathds{h}_{ij}}=-\mathds{h}_{ji}\left(\mathds{h}^{-1}\right)_{ii}\left(\mathds{h}^{-1}\right)_{jj}\\
+\frac{1}{2\lambda\tau}\sum_{k=1}^{3}\mathds{h}_{ik}(s-s')_{k}(s-s')_{j},
 \label{eq:19}
\end{split}
\end{equation}

The derivatives with respect to the potential action are
obtained in a similar manner, writing the function $u_{2}(r,r';\tau)$
in Eq.~(\ref{eq:13}) in terms of the matrix $\mathds{h}$ and the relative coordinates
$s$ and $s'$. In practice, it is useful~\cite{26} to express $u_{2}$ in terms of a
different coordinate set. Defining

\begin{equation}
\begin{split}
q\equiv\frac{1}{2}\left(\left|r\right|+\left|r'\right|\right)=\frac{1}{2}\left(\left|\mathds{h}s\right|+\left|\mathds{h}s'\right|\right),
 \label{eq:20}
\end{split}
\end{equation}

\begin{equation}
\begin{split}
z\equiv\left|r\right|-\left|r'\right|=\left|\mathds{h}s\right|-\left|\mathds{h}s'\right|,
 \label{eq:21}
\end{split}
\end{equation}

\begin{equation}
\begin{split}
t\equiv\left|r-r'\right|=\left|\mathds{h}\left(r-r'\right)\right|,
 \label{eq:22}
\end{split}
\end{equation}

the potential action then takes the form

\begin{equation}
\begin{split}
S_{\mathtt{pot}}=\sum_{n<m}^{N}u_{2}(q_{mn},z_{mn},t_{mn};\tau),
 \label{eq:23}
\end{split}
\end{equation}

where, for instance
\begin{equation}
\begin{split}
q_{mn}=\frac{1}{2}\left(\left|\mathds{h}(s_{n}-s_{m})\right|+\left|\mathds{h}(s_{n}'-s_{m}')\right|\right).
 \label{eq:24}
\end{split}
\end{equation}

The derivatives with respect to the h-matrix elements are
then given by

\begin{equation}
\begin{split}
\frac{\partial u_{2}(q,z,t;\tau)}{\partial H_{ij}}=\left(\frac{\partial q}{\partial H_{ij}}\frac{\partial u_{2}}{\partial q}+\frac{\partial z}{\partial H_{ij}}\frac{\partial u_{2}}{\partial z}+\frac{\partial t}{\partial H_{ij}}\frac{\partial u_{2}}{\partial t}\right)
 \label{eq:25}
\end{split}
\end{equation}

with
\begin{equation}
\begin{split}
\frac{\partial q}{\partial H_{ij}}=\frac{1}{2}\left(\frac{\sum_{k}H_{ik}s_{k}s_{j}}{\left|Hs\right|}+\frac{\sum_{k}H_{ik}s_{k}^{'}s_{j}^{'}}{\left|Hs'\right|}\right),
 \label{eq:26}
\end{split}
\end{equation}

\begin{equation}
\begin{split}
\frac{\partial z}{\partial H_{ij}}=\frac{1}{2}\left(\frac{\sum_{k}H_{ik}s_{k}s_{j}}{\left|Hs\right|}-\frac{\sum_{k}H_{ik}s_{k}^{'}s_{j}^{'}}{\left|Hs'\right|}\right),
 \label{eq:27}
\end{split}
\end{equation}

and 

\begin{equation}
\begin{split}
\frac{\partial t}{\partial H_{ij}}=\left(\frac{\sum_{k}H_{ik}\left(s-s'\right)_{k}\left(s-s'\right)_{j}}{\left|H\left(s-s'\right)\right|}\right).
 \label{eq:28}
\end{split}
\end{equation}

The calculation of the stress tensor is then based on using
expressions [19] and [25] in Eqs.~(\ref{eq:11})  and ~(\ref{eq:14}).

\section{III. COMPUTATIONAL DETAILS}
\label{sec:III. COMPUTATIONAL DETAILS}

All calculations are based on an orthorhombic computational
cell containing 180 $^{4}He$ atoms arranged on a hcp
lattice characterized by the ideal ratio $c/a = 1.633$, which
is within $0.3\% $ of the experimental estimates for all considered
densities~\cite{27}. The cell is constructed in such a manner that
the $x$, $y$, and $z$ axes are parallel to the crystallographic
$[\overline{1}\enskip 2\enskip \overline{1}\enskip 0]$, $[\overline{1} \enskip0\enskip 1\enskip 0]$, and  $[0\enskip 0\enskip 0\enskip 1]$ 
directions, respectively. Standard
periodic boundary conditions are applied throughout.
The simulations have been carried out using the $PIMC++$
package~\cite{18}, which is a $\mathrm{C++}$ implementation of the $PIMC$
algorithms described in Ref.~\cite{14}. The used pair action was
obtained from a standard matrix squaring procedure~\cite{14,15} using
the Aziz $\mathrm{HFD-B3-FCI1}$ pair potential~\cite{16} and an interaction
cutoff of 8$\AA$ . All $\mathrm{PIMC}$ simulations employ a time step
$\tau=\beta/m=1/40 K^{-1}$~\cite{28},  with $\beta$ the inverse temperature and
$M$ the number of beads in the ring polymers.
The determination of the elastic stiffness constants $C_{ijkl}$ is
based on the definition~\cite{29}.

\begin{equation}
\begin{split}
\sigma_{ij}=\sum_{k,l}C_{ijkl}\epsilon_{kl}
 \label{eq:29}
\end{split}
\end{equation}

where $\epsilon_{kl}$ is the $kl$ component of the strain tensor and $\sigma_{ij}$
is the $ij$ component of the corresponding stress tensor. The
stiffness constants are then determined by imposing different
kinds of strains 	$\epsilon_{kl}$  and measuring the induced stress responses
and $\sigma_{ij}$. Specifically, each simulation is carried out using a cell
in which only one of the six independent strain components
(measured with respect to the undeformed cell reference) is
different from zero. In this manner, there is only one term in
the summation of Eq.~(\ref{eq:29}). Exploring the linearity of Eq.~(\ref{eq:29}),
we determine the stress response as a function of the magnitude
of a given strain component. The stiffness constants are then
given by the slopes of the $\sigma_{ij}(\epsilon_{kl})$ graphs,

\begin{equation}
\begin{split}
C_{ijkl}=\frac{d\sigma_{ij}}{d\epsilon_{kl}}.
 \label{eq:30}
\end{split}
\end{equation}

\section{IV. RESULTS AND DISCUSSION}
\label{sec:IV. RESULTS AND DISCUSSION}

\begin{figure}
\begin{center}
\includegraphics[width=8.0cm]{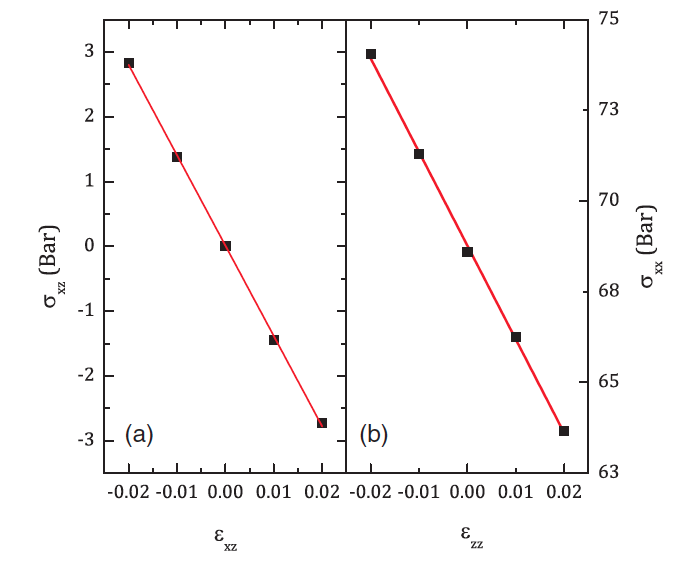}
\caption{(color online).  Stress response as a function of imposed
deformation. Magnitude of the error bars corresponds to size of
symbols. Full lines depict linear fits to the PIMC results. (a) Shear
stress $\sigma_{xz}$ in response to shear strain $\epsilon_{xz}$, scale of the left-hand side.
(b) Tensile stress $\sigma_{xx}$ a function of tensile strain $\epsilon_{zz}$, scale of the
right-hand side.}
\label{fig1}
\end{center}
\end{figure}

Figure~[\ref{fig1}] shows typical results for the stress components as
a function of the imposed strain at a temperature of 1 $\mathrm{K}$ and
a molar volume of 20 $\mathrm{cm}^{3}$. Figure~[\ref{fig1}-a] shows the shear stress
response $\sigma_{xz}$ as a function of the imposed shear strain $\epsilon_{xz}$ ,
where the indications $x$ and $z$ refer to the $x$ and $z$ directions
of the computational cell. In all cases, we verified that the
small distortions did not cause any disruptions of the crystal
structure, leading only to small homogeneous deformations of
the hexagonal structure within the elastic limit. The behaviour is
distinctly linear up to absolute strain values of $2\%$, as attested
by the linear fit shown by the full line. As expected, the line
passes through the origin, with no shear stress being present at
zero shear deformation. The slope of the line then determines
the shear elastic constant $C_{xzxz}$ or, in Voigt notation~\cite{29},
$C_{44}$, usually known as the shear modulus $\mu$. Figure~[\ref{fig1}-b] 
shows a similar response curve, plotting the tensile stress $\sigma_{xx}$ in
response to the tensile strain $\epsilon_{zz}$. In this case the stress response
does not pass through the origin given that the reference cell is
in a state of hydrostatic compression. Once again the behaviour
is manifestly linear, with the slope giving the value of the
elastic constant $C_{xxzz}$ or, in Voigt notation, $C_{13}$.

\begin{figure}
\begin{center}
\includegraphics[width=8.0cm]{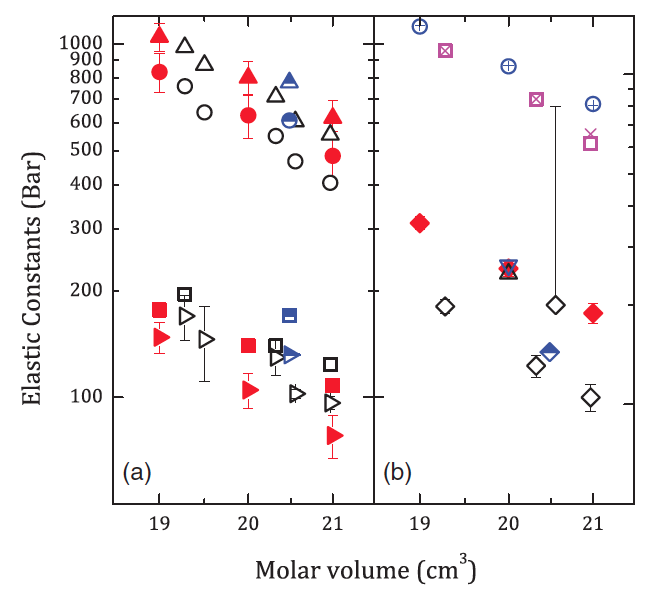}
\caption{(color online). (Color online) Elastic constants and some of their specific
combinations for hcp solid $^{4}$He as a function of the molar volume at a
temperature of 1K. (a) Filled symbols depict PIMC results. Half-filled
symbols represent zero-temperature VMC results of Ref.~\cite{22}. Open
symbols represent corresponding experimental data.When error bars
are not shown they are smaller than the symbol size. $C_{33}$ (upward
triangles), $C_{11}$ (circles), $C_{44}$ (squares), $C_{66}$ (right triangles). (b)
$C_{13}$ PIMC (filled diamonds, $\tau=1/40$K$^{-1}$, open upward triangle,
 $\tau=1/80$K$^{-1}$, open downward triangle, result obtained from bulk
modulus), $C_{13}$ experimental (open diamonds),$C_{13}$ VMC (half-filled
diamond), $C_{11}+C_{12}$ PIMC (o), $C_{33}+C_{13}$ PIMC (pluses), $C_{11}+C_{12}$
experimental (squares),$C_{33}+C_{1}$ experimental (crosses).}
\label{fig2}
\end{center}
\end{figure}

Using stress-strain curves of the type shown in Fig.~[\ref{fig1}] obtained from the PIMC simulations, we determine the 6 $×$ 6
elastic-constant matrix~\cite{29} of hcp solid $^{4}$He as a function of
the molar volume at a temperature of 1 $K$. As expected,
the results are found to obey the symmetry relations for
hexagonal crystals~\cite{29}, leaving only five independent stiffness
constants. These are shown in  Fig.~[\ref{fig2}], which also includes
sets of experimental data produced in the 1970s~\cite{30,35}. A
comparison between the theoretical and experimental data,
however, must be conducted with some care. Whereas the
PIMC calculations give values for the \textit{isothermal} stiffness
constants, the ultrasonic experimental data typically probe the
\textit{adiabatic} elastic constants. Accordingly, a direct comparison
between the two data sets is meaningful only if the difference
between these two kinds of elastic constants is small. In
its estimation, we applied the expressions for the difference
between the isothermal and adiabatic elastic constants~\cite{36} and
used experimental thermodynamic data~\cite{37} for the isochoric heat
capacity $C_{\nu}$ and the isochoric pressure coefficient $(\partial P /\partial V)_{\nu}$.
For a molar volume of 19.135 $cm^{3}$ and a temperature of 1$K$, we
find the difference to be of the order of $10^−{1}$ bar, which is very
small compared to the absolute values of the elastic constants
and their error bars, thus justifying a direct comparison of the
PIMC data with experiment.

Fig.~[\ref{fig2}-a] compares our PIMC results to experimental
data for the four elastic constants $C_{33}$, $C_{11}$, $C_{44}$, and $C_{66}$. For
further comparison we have also included results from recent variational Monte Carlo (VMC) for $T=0 K$ based on the
shadow wave function formalism~\cite{22}. The agreement between
PIMC and experiment is excellent for  $C_{33}$, and  $C_{11}$,, with values
essentially within each others error bars across the considered
density range. The agreement is also good for  $C_{44}$, and  $C_{66}$, with
the PIMC values $\sim$10 - 20\% below the experimental data. For
these four constants, the PIMC results also show slightly better
agreement with experiment compared to the VMC data, which
systematically overestimate all four constants by  $\sim$20 - 30\%.

The situation is markedly different for the remaining
independent elastic constant, $C_{13}$, however. As shown by
the diamond symbols in Fig.~[\ref{fig2}-b] the PIMC data overestimate
experiment by  50 - 100\%, with exception of the data
point measured by Greywall in 1971, although the latter is
characterized by an error bar of 300\%~\cite{32}. To further verify
our result we performed additional computations at a molar
volume of 20 cm$^{3}$. First, we carried out the PIMC calculations
using a reduced time step of $\tau = 1/80 K^{-1}$. Furthermore, we
also determined $C_{13}$ in an indirect way, computing the bulk
modulus $B = −V (\partial P/\partial V)_{T}$ by a finite-difference derivative
of the hydrostatic pressure with respect to volume changes
and using the relation $C_{13} =  (3B- C_{33})/2$. As can be seen
in Fig.~[\ref{fig2}-b], neither led to significantly different results for
$C_{13}$, lending further support to the internal consistency of our
calculations. Interestingly, the VMC result for $C_{13}$ is actually
in good agreement with experiment, differing by $\sim $10\%.
To investigate the origin of this discrepancy, it is useful
to verify whether certain relations, different from those associated with the crystal symmetry, hold. One of these is the relation

\begin{equation}
\begin{split}
C_{11}+C_{12}=C_{33}+C_{13},
 \label{eq:31}
\end{split}
\end{equation}

which should be satisfied in case the $c/a$ ratio is independent
of the density. This independence has been verified experimentally
for a wide range of pressure values~\cite{30}. This is consistent
with the fact that the experimental values of the left- and
right-hand sides of Eq.~(\ref{eq:31})  are essentially equal, as shown by
the open squares and crosses, respectively, in Fig.~[\ref{fig2}-b] .Despite
the overestimate for stiffness constant $C_{13}$, the PIMC results
are in fact consistent with this characteristic of solid $^{4}$He in
the hcp phase as shown by the open circles and pluses in the
same plot.

Another concerns the validity of the so-called Cauchy
relation~\cite{38,39}, $C_{13} - P = C_{44} + P$, where $P$ is the applied
hydrostatic pressure. These are relations between elastic
constants that are expected to hold when the (a) interatomic
interactions can be described by purely two-body central forces
and (b) when vibrational effects, zero-point or thermal in
origin, are negligible. It is useful to quantify deviations from
this relation through the parameter~\cite{35,36,37,38,39}

\begin{equation}
\begin{split}
\delta=\frac{C_{44}+P-\left(C_{13}-P\right)}{\left(C_{13}-P\right)}.
 \label{eq:31}
\end{split}
\end{equation}

For a classical crystal at $T=0$K and characterized by pairwise
central interaction forces one has $\delta=0$. Deviations from this
value are then a measure of the magnitude of vibrational effects
and the importance of many-body interaction forces. In the
case of quantum crystals at low temperatures, the deviation
thus is a probe for the role of zero-point motion and/or many body
interactions. It seems plausible, however, that the latter
are rather small for the condensed phases $^{4}$He~\cite{40}.

\begin{figure}
\begin{center}
\includegraphics[width=8.0cm]{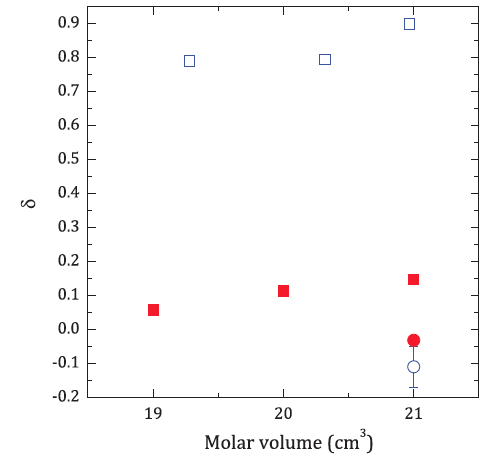}
\caption{(color online). Cauchy deviation values for the hcp
(squares) and bcc (circles) phases. Experimental results (open
symbols), PIMC results (full symbols).}
\label{fig3}
\end{center}
\end{figure}

Fig.~[\ref{fig3}]  shows $\delta$ as a function of the density for the
experimental data and our PIMC results for the hcp phase.
The results show a distinct discrepancy. While the calculations
give relatively small deviation values, ranging between 0.05
and 0.14, the experimental deviations are 0.8-0.9. Again, this
discrepancy can be traced back to the difference in the value
of stiffness constant $C_{13}$.

To further assess these issues we also computed the
elastic constants for the bcc phase, which exists at similar
densities and temperatures as those considered for the hcp
form. In this case, we find very good agreement between
PIMC and experiment for \textit{all} elastic constants. Specifically,
at a density of 21 cm$^{3}$ and a temperature of 1.5 $K$, the
PIMC calculations give $C_{11} = 371 \pm 43$, $C_{12} = 330 \pm 25$, and
$C_{44} = 217\pm 4$ bar. Inelastic neutron scattering measurements,
on the other hand, give $C_{11} = 349 \pm 15$, $C_{12} = 301 \pm 10$, and
$C_{44} = 215 \pm 8$  bar so that the differences between PIMC and
experiment are less than 10\% for all stiffness constants. As
a result, there are no significant discrepancies between PIMC
and experimental Cauchy deviations, as can be gauged by the
parameter $\delta = (C_{44} -C_{12} + 2P)/(C_{12}-P)$ in Fig.~[\ref{fig3}].

The interpretation of the above results is challenging. In the
present situation, our results agree very well with experiment
for four of the five independent elastic moduli of the hcp
phase, but show a deviation for the remaining constant, $C_{13}$. If
this were to be interpreted as a flaw in our calculations, then
discrepancies would also be expected for other phases. This
does not seem to be the case. In addition to previous results for
the liquid phase based on very similar interaction models~\cite{14}, our
calculations are in very good agreement with experiment for
all elastic constants of the bcc phase. A further element in this
discussion is that, although conducted for zero temperature,
recent VMC and diffusion Monte Carlo results~\cite{22,41} do not
seem to show this deviation for $C_{13}$ in hcp $^{4}$He.

To unveil this puzzle it would be extremely useful to
renew the experimental data for the elastic constants of hcp
$^{4}$He. The data sets available today are more than 30 years
old and, with present experimental capabilities, it should
be possible to obtain the elastic constants with significantly
greater accuracy. Not only would such data be useful toward
gauging current theoretical modeling approaches, they would
also be particularly useful, for instance, in understanding
the elastic properties of polycrystalline $^{4}$He samples~\cite{11}. In
addition, such information would also be of value in discriminating
between changes in elasticity and supersolidity,
respectively, when interpreting frequency changes in TO
experiments~\cite{11}.

\section{V. CONCLUSIONS}
\label{sec:V. CONCLUSIONS}

In summary, in this paper we have reported values of
the elastic stiffness constants of solid $^{4}$He in the hcp phase
determined using the PIMC approach based on the Aziz pair potential
model. To this end we have developed an expression
for the stress observable in the path-integral formalism,
allowing the direct measurement of the internal stress state of
a system in PIMC simulations. This development is of interest
in its own right, allowing for instance, an explicit atomistic
determination of the Peierls stress for dislocation motion~\cite{22}.
Here, we use it to compute the elastic stiffness constants by
measuring the linear stress response to imposed small strain
conditions.
Four of the five computed elastic stiffness constants as
a function of density show good agreement with experiment.
The stiffness coefficient $C_{13}$, which is $\sim$50-100\%
larger than reported experimental values, is an exception.
This discrepancy leads to very different deviations in the
Cauchy relation associated with $C_{44}$ and $C_{13}$. The same
calculations for the bcc phase, on the other hand, show
good agreement between experiment and PIMC for all elastic
constants.

\section*{ACKNOWLEDGMENTS}

We gratefully acknowledge support from the Brazilian
agencies CNPq, Fapesp, and Capes. We thank Prof. J. Beamish
for helpful discussions. The calculations were performed at
CCJDR-IFGW-UNICAMP and CENAPAD-SP.\\

$^{*}$ Corresponding author: dekoning@ifi.unicamp.br

\end{document}